\RequirePackage{fix-cm}
\documentclass[smallextended]{svjour3}       % onecolumn (second format)
\smartqed  % flush right qed marks, e.g. at end of proof
\usepackage{graphicx}
\newcommand{\pt}{$p_T$}

\begin{document}

\title{Iterative subtraction method for Feature Ranking%\thanks{Grants or other notes
}


\author{Paul Glaysher \and
        Judith M. Katzy \and 
        Sitong An}


\institute{DESY \at
              Notkestr.85, 22607 Hamburg, Germany \\
              Tel.: +49-40-89980\\
              \email{paul.glaysher@desy.de}\\
                            \email{judith.katzy@desy.de}           \\
                            \email{s.an@cern.ch}\\
                            All figures and pictures by the author(s) under a CC BY 4.0 license.\\
 }

\date{Received: date / Accepted: date}

\maketitle

\begin{abstract}
Training features used to analyse physical processes are often highly correlated and determining which ones are most important for the classification is a non-trivial tasks. For the use case of a search for a  top-quark pair produced in association with a Higgs boson decaying to bottom-quarks at the LHC, we compare feature ranking methods for a classification BDT. Ranking methods, such as the BDT Selection Frequency commonly used in High Energy Physics and the Permutational Performance, are compared with the computationally expense Iterative Addition and Iterative Removal procedures, while the latter was found to be the most performant. 


\end{abstract}

\section{Introduction}
\label{sec:intro}
Many measurements and searches for new phenomena performed by the experiments at the Large Hadron Collider (LHC)  use a Boosted Decision Tree (BDT) to discriminate the physics process of interest (signal)  from other physics processes with similar signature (background). The input variables (called features) to these BDTs are reconstructed from  detector signals at  different level of sophistication, hence forming low level and high level features. The variables are  usually chosen based on the understanding of the physical processes. The BDTs are typically trained by supervised learning on labeled events of simulated  signal and background processes using the Monte Carlo (MC) technique. The resulting trained BDT is  applied to unlabelled data to obtain measurement results.  \\

Knowing  the relative importance of the input variables, $i.e.$ ranking the features,   helps in various aspects.
Firstly, it allows  to reduce unnecessary dimensionality which is particularly  important when dealing with small training samples. This is often the case  when machine learning algorithms are used to classify physics processes which are CPU expensive to simulate and hence only a limited sample size exist for training and testing. Reducing dimensionality also  helps for faster training. For example, the runtime complexity, i.e. the CPU time needed to construct a  decision tree scales linearly with number of training variables \cite{witten}. While this may still be manageable for BDTs, experience shows that the training time for other machine learning algorithms (ML) such as neural networks may significantly increase with the number of input variables used. 

Feature ranking is also used as one possibility to gain insight into the underlying model of a physical process, i.e. the importance of the selected variables for the analysis. It allows for analysis optimisation such as validating the modelling of the inputs. Often potential training bias of the BDT response due to the particular MC generator used is estimated by using alternative MC simulations leading to a  slightly modified BDT response which is then propagated into the uncertainty of the measurements.  Feature ranking will lead to a better understanding of the source of this difference and help  reduce the measurement uncertainties. 

However, the question which training features are most important for the classification may not have a unique answer, in particular when the input variables are highly correlated.  Ranking variables to reduce dimensionality can be probed with training BDTs  on a sub-set  of variables with algorithms optimised to find the optimal sub-set. While the importance of a  variable for a given BDT classification might better be probed by using ranking algorithms estimating the effect of single variables  on the classification of the BDT trained with the full set of input variables.

This paper studies various existing and new algorithm to select the best variables to be used for training.

\section{Input variables and set-up}
\label{sec:setup}
The current study is inspired by the example of a classification BDT used in the search for the process of top-quark-pair production in association with a Higgs boson (ttH) performed by the ATLAS experiment at LHC \cite{ttHbb}. This search was performed in the Higgs decay channel where the Higgs decays to a pair of bottom ($b$) quarks. 
The signal events contain one electron or muon, at least six jets and 4 $b$-quark jets. The dominant background is top-quark pair production in association with a $b$-quark pair from gluon splitting which contains the same final state objects however, with slightly different kinematic properties. 

We use MC samples provided by the HepSim Group\cite{opendata}. The ttH signal sample containing $13\cdot 10^6$ events was generated with MadGraph \cite{Alwall:2011uj} matched to the Herwig6 parton shower \cite{Corcella_2001}. Two background samples were generated: $2\cdot 10^6$ events of top-pair production with additional light quarks  using MadGraph  matched to the Herwig6  and $10\cdot 10^6$ events of top-pair production with additional $b$-quarks  using MadGraph matched to Pythia6\cite{Sjostrand:2006za} . The two background samples are orthogonal and are merged into one background sample  % "tt+jets" sample
 with the different processes weighted by their cross section. The ATLAS detector response  was simulated using Delphes simulation \cite{delphes}.  For this study, reconstructed jets and $b$-quark jets (called $b$-jets in the following) are used. The reconstructed $b$-jets have a 70\% tagging probability. The corresponding light jet/c-jet rejection probability is parameterised according to \cite{Aaboud:2018xwy}.

Events selected for the BDT training were required to fulfil the following criteria:
\begin{itemize}
\item one electron or muon with transverse momentum \pt $\geq$ 20\,GeV 
\item at least 5 jets with \pt $\geq$ 25\,GeV
\item at least 3 $b$-jets.
\end{itemize}

After this selection 700 000 signal events and 275000 background events remain. From these events two thirds are used for training a BDT and one third to test the BDT. 

\begin{figure}
\includegraphics[width=1.0\textwidth]{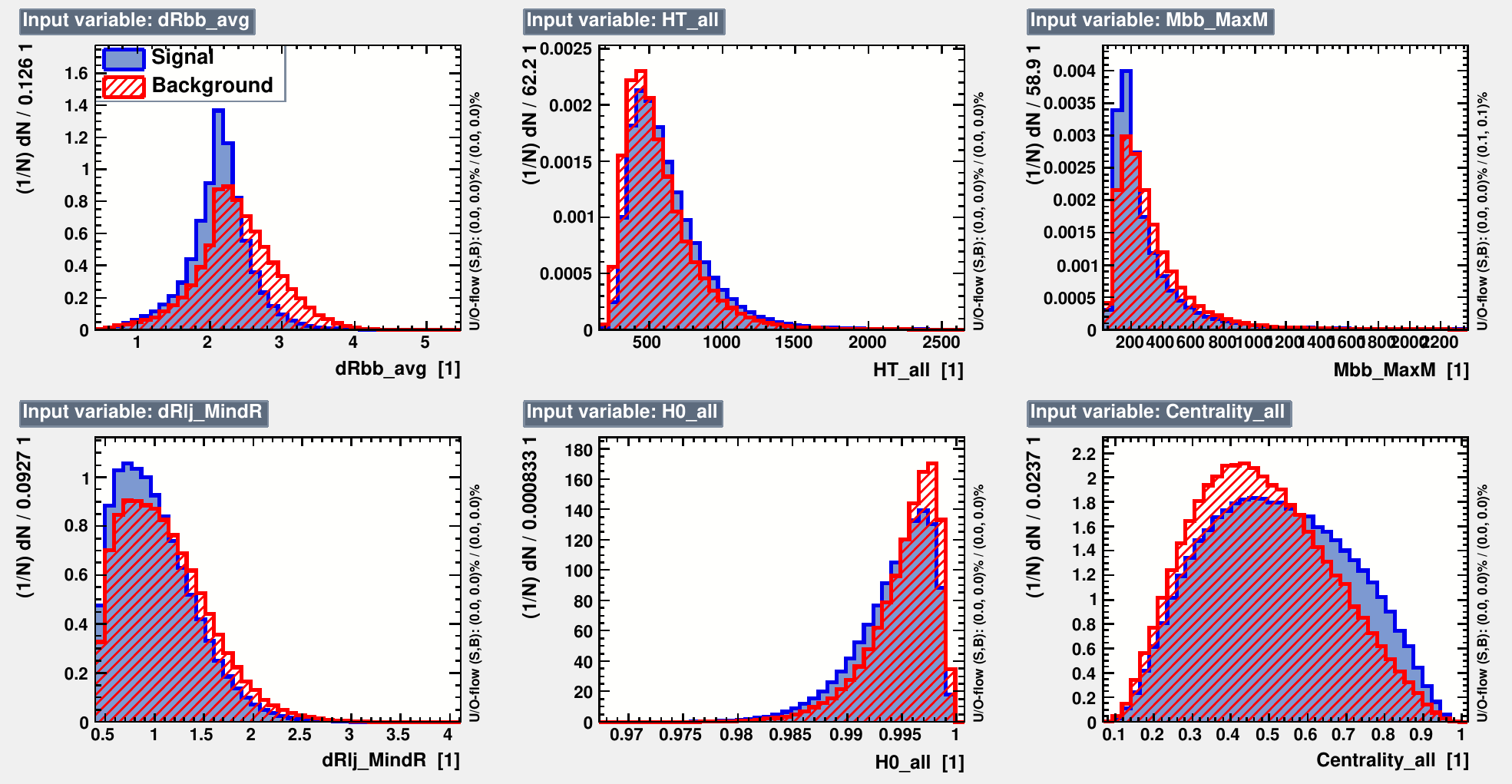}
\caption{Input variables to the BDT from signal (blue) and background (red) samples, for variable definition see text.}
\label{fig:input}       % Give a unique label
\end{figure}

\begin{figure}
\includegraphics[angle=-90, width=1.0\textwidth]{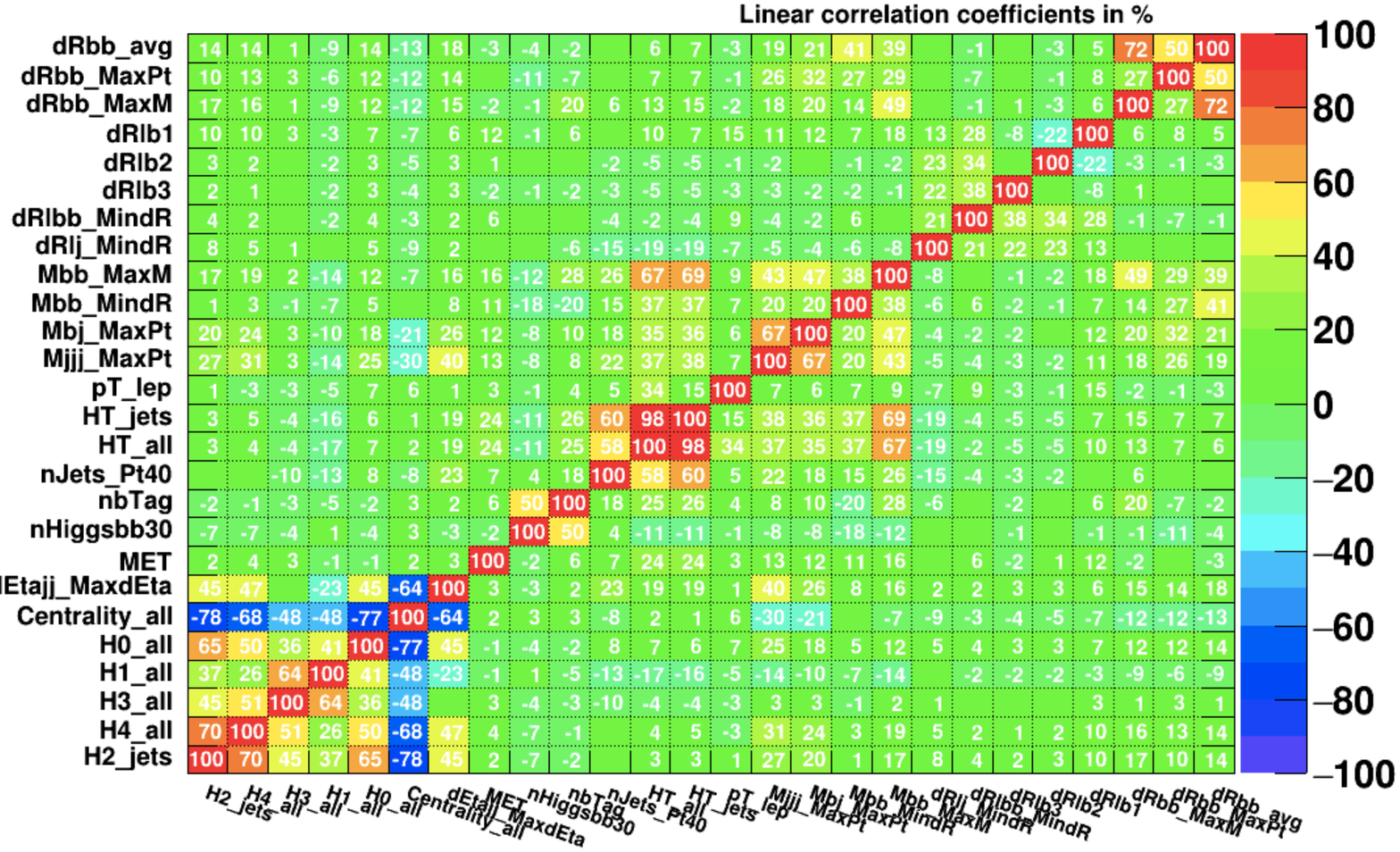}
\caption{Matrix of linear correlation coefficents of the  input variables to the BDT.}
\label{fig:correlation}       % Give a unique label
\end{figure}

The choice of  training variables is inspired by the reference analysis \cite{ttHbb}, with a few additional variables and removing variables that could not easily be reconstructed from the available information.  In total 26 input variables are considered ranging from  basic objects like angular distance $dR$ between different jets or leptons,  mass of various jet and/or lepton systems, scalar sum of the \pt of jets and leptons and the  full event topology. The complete list of variables is given in Tab.\ref{tab:inputs}.
%39 variables are computed from which 26 are considered. These variables were selected based on the requirement to have at least 1\% separation in signal versus background shape. \TODO: how is separation defined? Difference in integral? Add to TMVA histogramms, can't we start with 39 and use TMVA ranking first?)
Figure \ref{fig:input} shows distributions of input variables in the signal and the background sample. The separation, defined as the integral over the absolut value of the difference between signal and background, varies between 1\% and 8 \%.  
  Figure \ref{fig:correlation} shows the correlation of the variables, ranging from almost no correlation to very high (anti-) correlation.

The TMVA \cite{tmva} implementation of the BDT code is used with 400 trees, a maximal depth ("MaxDepth") of 5 and the Ada boosting algorithm ("AdaBoostBeta=0.15) and 80 Cuts ("nCuts=80").

\begin{table}[htb]
\begin{tabular}{|  l | l| }
\hline
dRbb\_avg & average dR of all $b$-jet pairs \\
dRbb\_MaxPt & dR of the $b$-jet pair with the highest sum of \pt \\
dRbb\_MaxM & dR of the $b$-jet pair with the highest invariant mass \\
dRlb1-dRlb3 & dR of the charged lepton and the $b$-jet with the 1st-3rd largest \pt \\
%dRlb2 & dR of the charged lepton and the $b$-jet with the 2nd largest \pt \\
%dRlb3& dR of the charged lepton and the $b$-jet with the 3rd largest \pt \\
dRlbb\_MindR & dR of the charged lepton and total $b$-jet pair system which \\
& has the smallest dR \\
dRlj\_MindR & minimum dR between the charged lepton and any jet \\
Mbb\_MaxM & maximum invariant mass of any $b$-jet pair \\
Mbb\_MindR & invariant mass of $b$-jet pair which has the smallest dR \\
Mbj\_MaxPt & invariant mass of two jets with the largest \pt \,sum, \\
& where exactly one of the jets is a  $b$-jet\\
Mjjj\_MaxPt & invariant mass of any three jets with the largest \pt sum \\
pT\_lep & transverse momentum of the charged lepton \\
HT\_jets & sum of transverse momentum of all jets \\
HT\_all & sum of transverse momentum of all jets and the charged lepton \\
nJets\_Pt40 & number of jets with \pt$ \geq 40$ GeV \\
nbTag & number of $b$-jets \\
nHiggsbb30 & number of $b$-jet pairs with an invariant mass\\
 &  within 30 GeV of the Higgs boson mass of 125 GeV \\
MET & missing transverse energy \\
dEtajj\_MaxdEta & largest difference in longitudinal angle $\eta$ of any two jets \\
Centrality\_all & ratio of momentum sum over the energy sum of all objects \\
H$i$\_all, H2\_jets & 1st-5th Fox Wolfram transverse moment \cite{fox} of all objects \\
%H1\_all & 2nd Fox Wolfram transverse moment of all objects \\
%H2\_jets & 3rd Fox Wolfram transverse moment of all jets \\
%H3\_all & 4th Fox Wolfram transverse moment of all objects \\
%H4\_all & 5th Fox Wolfram transverse moment of all objects \\
 \hline
\end{tabular}
\caption{\label{tab:inputs} Input variables used for the BDT.}
\end{table}

In the following,  the feature ranking  of the $N=26$ input variables  are compared using different methods. For each method, BDTs are trained for the set of  the $n$ highest ranked  variables and the area under the receiver operation curve (AUROC) is taken as the performance measure for comparison. The difference in ranking may lead to a different sub-set of variables from the full set of variables for a given fixed number  $n$ of variables. However, for each method the list of variables used builds up sequentially for each algorithm, i.e. exact one variable is added to the existing sample going from $n$ to $n+1$ variables and hence defines the ranking.  Only for the random selection the sub-samples for different  number of variables don't have to have any overlap as at each $n$ the random selection computed from all permutations for $n \leq 3$, and for 1000 randomly selected trials for $n>3$.

\section{Feature ranking algorithms}
\label{sec:algorithm}
Different algorithms for  ranking the importance of a feature (i.e. input variable) exist which largely vary in their methods. Some methods evaluate the variable importance by adding or subtracting input variables from or to a set of reference variables and measure the change in BDT performance. Other methods estimate the importance for a given set of variables based on the information used in the training of the BDT trained on all $N$ features.  The choice of method may also depend on the particular use case. The methods vary largely in their computing needs, some are very computationally expensive.\\

%\begin{itemize}

$\mathbf{Random\, Selection}$  For the first $n = 3$ variables, all $\frac{N!}{n! (N-n)!}$ possible combinations are considered and the one with the best AUROC is selected ("maximum").  This corresponds to the best possible AUROC for the given number of variables. However, since the number of combinations raises fast,  for $n\geq 4$ only a  random selection out of all combinations is choosen 1000 times, to limit CPU consumption on the BDT trainings. The median and the best  AUROC of all trails is reported to serve as a reference. \\

 $\mathbf{ Separation \,based}$ Rank variables by overlap of their signal versus background predictions, i.e. the integral over their difference. This method does not involve any BDT training. For the comparisons presented here, the AUROC values are calculated from a BDT trained with the $n$ selected variables.\\

 $\mathbf{Correlation \, based}$ Rank the variables based on their correlation to the BDT score computed with all variables. Computationally cheap as it only involves only one  BDT training with all $N$ variables. \\

$\mathbf{BDT \, Selection\, Frequency }$ Train the BDT on all $N$ variables and rank by how often a variable provided the optimal decision in the BDT \cite{tmva}. Computationally cheap as it  involves only one BDT training. This is the default ranking procedure implemented in the TMVA BDT code used here.\\

$\mathbf{Permutational \, Performance\, or \, Mean\, Decrease\, Accuracy\, (MDA)}$ In order to avoid the high CPU costs of the iterative removal and gives insight into a black-box estimator for the set of $N$ variables,  this method calculates the feature importance by replacing a feature with random noise instead of removing the feature. The random noise is  drawn from the same distribution as original feature values but taken from other events feature values. This avoids out of range values which may lead to a failure of the algorithm but the values are random and uncorrelated to the events ~\cite{permperf}\\

 $\mathbf{ Iterative \,Addition}$ The idea is to  measure the importance by looking at how much the score increases when a feature is added. It
 starts with the single input variable with highest AUROC and successively adds the variable of the remaining $N-n$  variables with the highest AUROC, as is done e.g. in \cite{Aaboud:2018psm}. 
This involves training the BDT for each of the $N-n$ combinations to determine the AUROC and find the best performance. The total number of BDTs to be trained are $\sum_{n=0}^N{(N-n)} = N \cdot (N - 1)/2 = 406$. However this ignores potential correlations between the added variables. For example, two correlated variables might only provide separation power when both are present in the training.\\

%Ideally all combinations to add n out of N variables should be tested, leading to $\frac{N!}{n! (N-n)!}$ combinations. But this could lead to series where e.g. the highest ranked variable is not part of the highest 3 ranked variables etc..

 $\mathbf{Iterative \, Removal}$ The idea is to  measure the feature importance by looking at how much the score decreases when a feature is removed. This way, correlations between variables are better taken into account than for the additive method. However, since the set of $n$ variables is retrained,  it shows what may be important within the dataset, not necessarily what is important within a concrete trained model.

This method starts with training on all $N$ variables and successively remove the variable that degrades the performance the least. As for the iterative addition, this involves training the BDT for each of the $ N-n$ combinations to determine the AUROC and find the best performance, leading in total to the same number of trainings as for the iterative addition. However, since the method starts with a larger number of variables in the BDT, the overall CPU consumption is high and even higher than the iterative addition. The code is publicly available \cite{code}.\\

%\end{itemize}

\section{Results}
\label{sec:results}

\begin{figure}
  \includegraphics[width=1.0\textwidth]{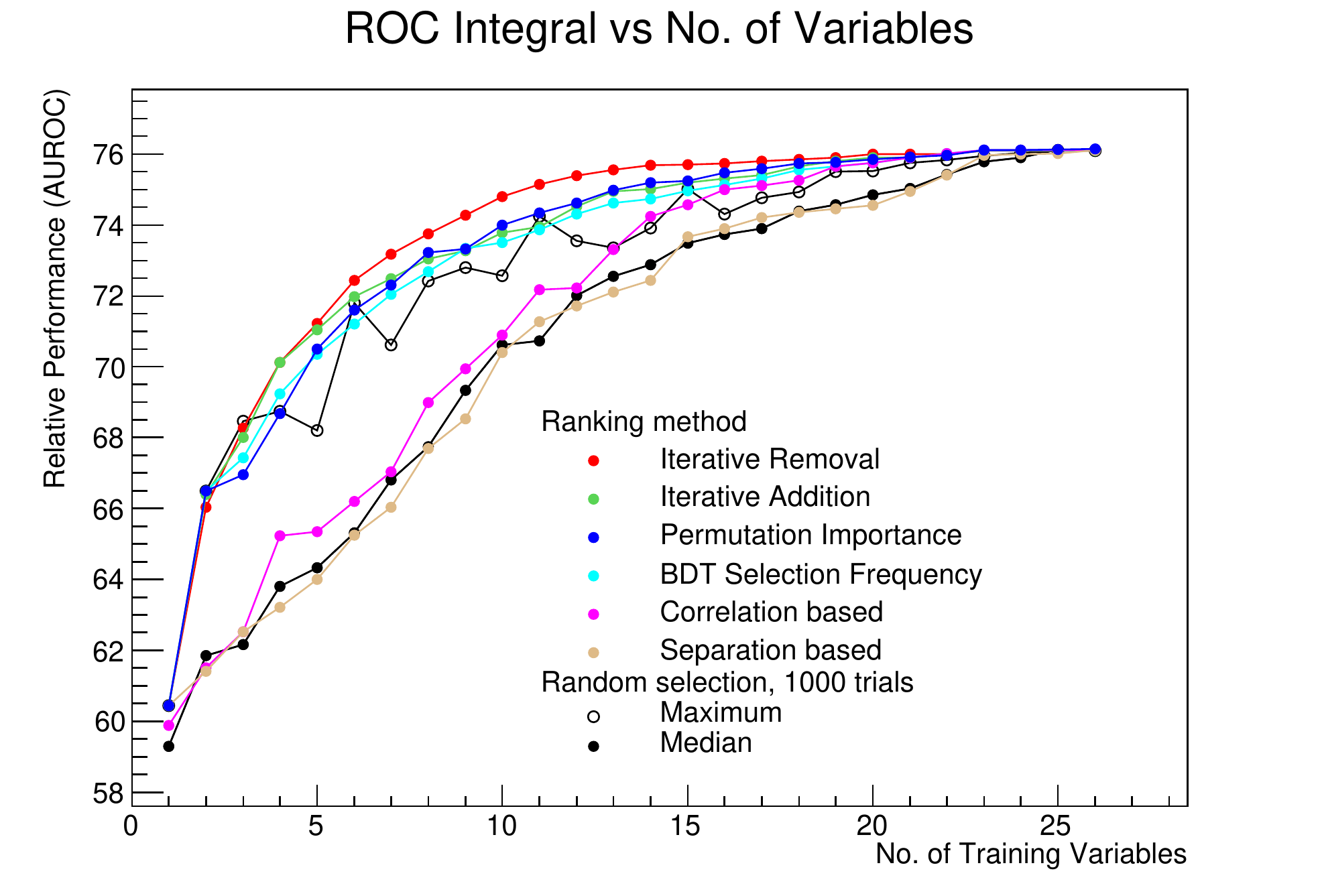}
\caption{Comparison of performance of different feature ranking algorithms. The area under the receiver-operater curve (AUROC) is shown as a function of the number of variables. For details of the selected number of training variables, see text.}
\label{fig:auroc}       % Give a unique label
\end{figure}

\begin{table}[htb]
\begin{tabular}{| l| l| l| l| l|}
\hline
Rank & Iterative Removal & Permutation Importance & BDT Selection & Best \\
 & & & Frequency  & \\
\hline
1 & dRbb\_av & dRbb\_av & dRbb\_av & dRbb\_av \\
2 & HT\_jets & HT\_all & Mbb\_MaxM & Mbb\_MaxM \\
3 & nHiggsbb30 & H0\_all & HT\_jets & nHiggsbb30\\
4 & Mbb\_MaxM  & Mbb\_MindR & H0\_all & - \\
5 & nbTag & dRlj\_MindR & nJets\_Pt40 & - \\
6 & Mbb\_MinR & Mbb\_MaxM & dRlb2 & - \\
7 & dRlb3 & Centrality\_all & Mjjj\_MaxPt & - \\
8 & H2\_jets & Mbj\_MaxPt & pT\_lep & - \\
9 & H0\_all & HT\_jets & dEtajj\_MaxdEta & - \\
10 & Mjjj\_MaxPt & H2\_jets & dRlb1 & - \\
\hline
\end{tabular}
\caption{\label{tab:rank} Highest ranked variables for Iterative Removal, Permutation Importance, BDT Selection Frequency and  best combination for up to 3 input variables. Even though the best combination is determined considering all combinations for each number of variables, the resulting best combination included the previously ranked variables.}
\end{table}

All algorithms show similar performance if 24 out of 26 variables are used. However, 
the different algorithms  approach this plateau of maximal AUROC differently as a function of  number of variables. There are two groups of algorithms with similar performance:
The first group of algorithms are the Iterative Removal, Iterative Addition,  the BDT Selection Frequency and the Permutation Importance. These algorithms start with  a higher AUROC and approach the plateau faster. Among these the newly proposed Iterative Removal performs best over the full range and approaches the plateau up to a level of 99\%  already with 12 variables. 
This is better than the similar algorithm of iterative addition and the Permutational importance which reach this performance only with 16 variables, or the BDT selection frequency which needs 17 variables. Between these methods the largest differences are observed between 5 and 16 variables. The better performance of the iterative removal comes at high CPU costs and it is interesting to note that Permutation importance which is computational cheap has the 2nd best performance overall and yields similar good results as iterative removal for more than 16 variables. The BDT selection frequency which is  also computational cheap is only slightly worse than the Permutation importance.

 The second group consists of the Median of the Random Selection, separation based and correlation based selections start with a low AUROC and only slowly approach the plateau with 24 variables. Among these algorithms, the separation based is the poorest ranking method over the full range as might be expected since this method ignores the correlation between the variables. The correlation based outperforms the random choice when approaching the plateau and has similar performance at low number of variables.

The Maximum of the Random Selection apparently largely depends on the randomly selected variables for 4 up to approximately 18 variables. 1000 trials is not enough to approximate the best result, and the dependence on the particular selection of variables is still large, hence more variables sometimes yield a smaller AUROC.  The better reference is the median which shows steady raising AUROC. It is interesting to note that the separation based selection is lower than the median of the Random Selection for almost all numbers of variables below $n$=15, indicating that separation is a not a good quantity for feature ranking in contrast to intuition.

When selecting the highest ranked variables to reduce dimensionality it is interesting to know how much the different ranking algorithms overlap. Table\,\ref{tab:rank} lists the highest ranked variables for the best performing algorithms. It is worth noting that they all agree for the highest ranked variable, but largely vary in the order of the rest of the variables. The variations persist 
in the best 5 variables, but the Iterative Removal and the Permutation Importance have 6 variables in common among the best 10. Two out of  the 4 variables that were not included in  Permutation Importance  have high linear  correlation coefficients with variables that were in the list of Iterative Removal, there is a  large overlap for 8 out of 10 variables.  Similar conclusions hold for the comparison of BDT Selection Frequency with the Iterative 
Removal. 

%Depending on the use case - in particular whether reduction of dimensionality or influence of a single variable on the classification result - the 

\section{Conclusion}
Different methods for ranking input variables for BDT classification were compared. The computationally most expensive method of Iterative Removal showed the best classification power measured in terms of AUROC. However, when selecting 16 out of 26 variables, other methods such as Permutation Importance and BDT Selection Frequency which are computationally very cheap give very similar results. Interestingly these  methods select the same variable as being best.  Difference in performance between 2 and 16 variables for the 3 methods are of the order of 1-2\% in AUROC for the same number of variables and should be compared to the computational costs for the specific use case.
The  variables selected to be the best 10 agree to large extend between these two methods which may allow for reducing dimensionality with the less CPU expensive method. But the best 5 variables vary largely, which will make it difficult to draw firm conclusions on the impact of first few variables on the classification power.

%Text with citations \cite{RefB} and \cite{RefJ}.
%\subsection{Subsection title}
%\label{sec:2}
%as required. Don't forget to give each section
%and subsection a unique label (see Sect.~\ref{sec:1}).
%\paragraph{Paragraph headings} Use paragraph headings as needed.
%\begin{equation}
%\end{equation}

%\begin{table}
%\begin{tabular}{lll}
%\hline\noalign{\smallskip}
%\end{tabular}
%\end{table}
%

%\begin{acknowledgements}
%\end{acknowledgements}



\begin{thebibliography}{}

\bibitem{witten}
I. H. Witten, E. Frank and M. A. Hall\\
"Data Mining : Practical Machine Learning Tools and Techniques"\\
 Elsevier Science, 2011.

  \bibitem{ttHbb}
  M.~Aaboud {\it et al.} [ATLAS Collaboration],
  %``Search for the standard model Higgs boson produced in association with top quarks and decaying into a $b\bar{b}$ pair in $pp$ collisions at $\sqrt{s}$ = 13  TeV with the ATLAS detector,''
  Phys.\ Rev.\ D {\bf 97} (2018) no.7,  072016
  doi:10.1103/PhysRevD.97.072016
  [arXiv:1712.08895 [hep-ex]].

 
  \bibitem{opendata}
  S.~V.~Chekanov [HepSim Group],
  [arxiv:1403.1886 [hep-ph]].
  %  tev13pp\_mg5\_ttbar\_jet\_MadGraph/HW6\\
  %tev13pp\_mg5\_ttbar\_bjet\_MadGraph/P6\\
 % tev13pp\_mg5\_ttH\_MadGraph/HW6
 
   
  \bibitem{Alwall:2011uj}
  J.~Alwall, M.~Herquet, F.~Maltoni, O.~Mattelaer and T.~Stelzer,
  %``MadGraph 5 : Going Beyond,''
  JHEP {\bf 1106} (2011) 128
  doi:10.1007/JHEP06(2011)128
  [arXiv:1106.0522 [hep-ph]].

    
 \bibitem{Corcella_2001}
 G.Corcella,  I. G Knowles, G. Marchesini,  S. Moretti,  K. Odagiri,  P. Richardson, M. H .Seymour,  B. R. Webber
 JHEP {\bf 2001}(2001)010,
 doi:10.1088/1126-6708/2001/01/010

  
\bibitem{Sjostrand:2006za}
  T.~Sjostrand, S.~Mrenna and P.~Z.~Skands,
  %``PYTHIA 6.4 Physics and Manual,''
  JHEP {\bf 0605} (2006) 026
  doi:10.1088/1126-6708/2006/05/026
  [hep-ph/0603175].

 
  \bibitem{delphes}
  J.~de Favereau {\it et al.} [DELPHES 3 Collaboration],
  %``DELPHES 3, A modular framework for fast simulation of a generic collider experiment,''
  JHEP {\bf 1402} (2014) 057
  doi:10.1007/JHEP02(2014)057
  [arXiv:1307.6346 [hep-ex]].
    
 \bibitem{Aaboud:2018xwy}
  M.~Aaboud {\it et al.} [ATLAS Collaboration],
  %``Measurements of b-jet tagging efficiency with the ATLAS detector using $ t\overline{t} $ events at $ \sqrt{s}=13 $ TeV,''
  JHEP {\bf 1808} (2018) 089
  doi:10.1007/JHEP08(2018)089
  [arXiv:1805.01845 [hep-ex]].
  
  \bibitem{fox}
  %Fox Wolfram moments Ref: 
  G.C.~Fox and S.~Wolfram, 
  Nucl. Phys. B149 (1979) 413.
  
\bibitem{tmva}
A.Hoecker et al.\\
 TMVA - Toolkit for Multivariate Data Analysis 
[arXiv:physics/0703039]

 \bibitem{permperf}
   M. Korobov, K. Lopuhin \\
  Revision 371b402a.\\
  https://eli5.readthedocs.io/en/latest/blackbox/permutation\_importance.html

  
\bibitem{Aaboud:2018psm}
  M.~Aaboud {\it et al.} [ATLAS Collaboration],
  %``Performance of top-quark and $W$-boson tagging with ATLAS in Run 2 of the LHC,''
  Eur.\ Phys.\ J.\ C {\bf 79} (2019) no.5,  375
  doi:10.1140/epjc/s10052-019-6847-8
  [arXiv:1808.07858 [hep-ex]].

\bibitem{code}
S.An \\
Github vSearch repository:
 A parallelised script for variable selection and  the iterative removal method\\
https://github.com/sitongan/vSearch

% etc
\end{thebibliography}


\end{document}